\documentclass[aps,prd,preprint,groupedaddress,showpacs]{revtex4}

\usepackage{graphicx}
\usepackage{dcolumn}
\usepackage{bm}

\begin{document}

\preprint{hep-ph/0203274}

\title{Constraints on $\tan{\beta}$ in SUSY SU(5) GUT}


\author{Jun Tabei}
 \email{jun@hep.phys.waseda.ac.jp}
\affiliation{%
Department of Physics, Waseda University, Tokyo 169, Japan
}%

\author{Hiroshi Hotta}
 \email{hotta@hep.phys.waseda.ac.jp}
\affiliation{
Institute of Material Science and Technology, Waseda University, 
Tokyo 169, Japan
}%


\date{\today}

\begin{abstract}
On the evaluation of $\tan{\beta}$ is revisited 
by the method of the renormalization group 
with the criteria of the bottom-tau unification 
in the SUSY SU(5) GUT model. 
Among the conventional supersymmetric theories, 
an energy-scale constant $M_{SUSY}$ is introduced by-hand 
to decouple the lower energy-scale region 
from the effects of the supersymmetric particles. 
Provided that $M_{SUSY}$ really exists, 
only small $\tan{\beta} \simeq 2$ is allowed theoretically. 
While, medium value as $\tan{\beta} = 6 \sim 25$ 
is also supported if $M_{SUSY}$ is vanished away from the model. 
\end{abstract}


\maketitle

\section{Introduction}               
\begin{itemize}
\item[1)] - What are unified ? \\
As one of the general features of the SU(5) GUT models, 
both Yukawa coupling constants of the bottom-quark 
and the tau-lepton are unified at GUT scale. 
Additionally, in several SU(5) models including 
the {\em SUSY} SU(5) GUT model\cite{Nilles}, 
the evidence for the unification of gauge coupling constants 
is numerically provided\cite{Amaldi} . 

\item[2)] - What is $M_{SUSY}$ ? \\
In the conventional analysis\cite{Amaldi}, 
the gauge unification in the SUSY SU(5) GUT model 
is achieved by introducing 
an energy-scale constant $M_{SUSY}$ by-hand 
to decouple the lower energy-scale region 
from the effects of the supersymmetric particles. 
Thus, $M_{SUSY}$ is regarded as the threshold 
between the standard model(SM) and SUSY. 
The value of $M_{SUSY}$ 
depends on the strong coupling constant $\alpha_{3}$ 
very sensitively\cite{Amaldi}. 
For instance, while $M_{SUSY} \simeq$ 1(TeV) for 
$\alpha_{3}=0.106$ is given\cite{Amaldi} in 1991, 
$M_{SUSY} \leq$ 100(GeV) is obtained 
for the recent value\cite{Particle} of $\alpha_{3}=0.1185$. 
If $M_{SUSY}$ were really less than 100(GeV), 
we should have already detected such supersymmetric particles 
in the region of $\sim$ 100(GeV) experimentally, however 
we have not yet. This fact may allow us to think 
of $M_{SUSY}$ as a negligible parameter, or, at least, 
the authors should discuss on the possibility of the 
models without $M_{SUSY}$. 
Note that such model without $M_{SUSY}$ means the pure 
supersymmetric theory without including the conventional 
standard model. 

\item[3)] How much $\tan{\beta}$ is optimal ? \\
$\tan{\beta}$ is evaluated by the two different ways of 
the likelihood analysis at two-loop level in this paper. 
One of the two ways is 
on the unification of the gauge coupling constants, 
and another is of the bottom-tau's Yukawa coupling constants. 
The sequence to optimize $\tan{\beta}$ 
is as follows:\\
In the first, $M_{SUSY}$ and the GUT scale $M_{X}$ are 
fixed by the $\chi^2$-fitting on the unification 
of the gauge$(\chi^{2}_{g})$ coupling constants. 
Once $M_{SUSY}$ and $M_{X}$ are fixed, corresponding 
$\tan{\beta}$ is derived numerically to realize 
the bottom-tau unification at $M_{X}$ scale. \\ 
Additionally, the analysis on the model without $M_{SUSY}$ 
is also carried out in the same manner. 

\end{itemize}

\section{Analysis of $\tan{\beta}$ with $M_{SUSY}$}

On the optimization of $\tan{\beta}$ 
is discussed\cite{Barger} 
by the renormalization group equations(RGEs) 
of the gauge and Yukawa 
coupling constants at two-loop level\cite{Arason,Castano} 
in the SM and SUSY. 
As the Yukawa coupling constants, heavier three 
of the ordinal Fermions 
(top-quark, bottom-quark, and tau-lepton) 
are taken into consideration. \\
First of all, the explicit values of the parameters 
in this paper are referred from the latest issue 
of the Particle Data Group\cite{Particle}. \\
On the boundary conditions of the RGEs, they are described 
further below. 
The conditions of the gauge coupling constants 
are given on Z-boson's mass shell $M_{Z}=91.184$(GeV) 
in the $\overline{\mbox{MS}}$-scheme as: 
\begin{eqnarray}
\alpha_{1}
&=&
\frac{5}{3} \ 
\frac{\alpha_{\overline{MS}}}{1-\sin^{2}{\theta}_{\overline{MS}}}
\ , \\
\alpha_{2}
&=&
\frac{\alpha_{\overline{MS}}}{\sin^{2}{\theta}_{\overline{MS}}}
\ , 
\end{eqnarray}
as their explicit values are: 
\begin{eqnarray}
\alpha_{\overline{MS}}^{-1}&=&127.934 \pm 0.027
\ , \label{eq:ninva} \\
\sin^{2}{\theta}_{\overline{MS}}&=&0.23117 \pm 0.00016
\ , \label{eq:nsin2} \\
\alpha_{3}&=&0.1185 \pm 0.0020
\ . \label{eq:na3} 
\end{eqnarray}
Moreover, the definition of $\tan{\beta}$ at $M_Z$ is 
given as usual: 
\begin{equation}
\tan{\beta}(M_{Z})
\doteq
v_{2}(M_{Z})/v_{1}(M_{Z}) 
\ , \label{eq:tanb} 
\end{equation}
where, $v_1$ and $v_2$ imply the vacuum expectation values(VEVs) 
of two Higgs doublets, and as their explicit value: 
\begin{equation}
v(M_{Z})
=
(v_{1}^{2}(M_{Z})+v_{2}^{2}(M_{Z}))^{1/2}=246\mbox{(GeV)}
\ , \label{eq:nv} 
\end{equation}
is fixed. For simplicity, mere $\tan{\beta}$ 
means $\tan{\beta}(M_{Z})$ in this paper. \\
On the Higgs quartic coupling constant $\lambda$, this is 
precisely defined at $M_{SUSY}$\cite{Barger} as: 
\begin{equation}
\lambda(M_{SUSY})=
\pi 
\left[ \frac{3}{5}\alpha_{1}(M_{SUSY}) + \alpha_{2}(M_{SUSY}) \right] 
\cos^{2}{2\beta}(M_{SUSY})
\ , \label{eq:lmsusy}
\end{equation}
however, it is very difficult to derive $\lambda$ 
from this definition. As $\lambda$, the authors make 
use of an approximate, and easier alternative definition 
at $M_{Z}$: 
\begin{equation}
\lambda 
\doteq 
\pi 
\left[ \frac{3}{5}\alpha_{1}(M_{Z}) + \alpha_{2}(M_{Z}) \right]
\cos^{2}{2\beta}(M_{Z})
\ . \label{eq:lmz}
\end{equation}
Note that the numerical difference between 
the definitions Eq.~(\ref{eq:lmsusy}) and Eq.~(\ref{eq:lmz}) 
is very small as about $0.1\%$, 
and $\lambda$ is decoupled to the Yukawa coupling constants 
among the RGEs at one-loop level, 
therefore, all the numerical results in this paper 
are effectually independent of the choice 
of the definition of $\lambda$. \\
Boundary conditions of the Yukawa coupling constants 
are given at each on-shell energy-scales as: 
\begin{equation}
\alpha_{q}(m_{q})
=
\frac{m_{q}^{2}(m_{q})}{2 \pi v^{2}(m_{q})}
\ , 
\end{equation}
where, the suffix $q$ implies the flavor of Fermions 
as $q=t$, $b$, or $\tau$, respectively. Their explicit values 
are as follows: 
\begin{eqnarray}
m_{t}(m_{t})&=&174.3 \pm 5.1 \mbox{(GeV)} \ , \\
m_{b}(m_{b})&=&4.2 \pm 0.2 \mbox{(GeV)} \ , \\
m_{\tau}(m_{\tau})&=&1.77699 \pm 0.00029 \mbox{(GeV)} 
\ . 
\end{eqnarray}
When assuming the existence of $M_{SUSY}$, corresponding 
boundary conditions are given as: 
\begin{eqnarray}
\alpha^{SUSY}_{i}(M_{SUSY})&=&\alpha^{SM}_{i}(M_{SUSY}) 
\ , \ (i=1, 2, 3) 
\ , \\
\ &\ & \nonumber \\
\alpha^{SUSY}_{t}(M_{SUSY})
 &=&\frac{m^{2}_{t}(M_{SUSY})}
{2 \pi v^{2}(M_{SUSY})\sin^{2}{\beta}(M_{SUSY})}
\  \\
 &=& \alpha^{SM}_{t}(M_{SUSY})\frac{1}{\sin^{2}{\beta}(M_{SUSY})}
\  \nonumber \\
 &=& \alpha'^{SM}_{t}(M_{SUSY})
\ , \nonumber \\
\ &\ & \nonumber \\
\alpha^{SUSY}_{b,\tau}(M_{SUSY})
 &=&
\alpha^{SM}_{b,\tau}(M_{SUSY})\frac{1}{\cos^{2}{\beta}(M_{SUSY})}
\  \\
 &=& \alpha'^{SM}_{b,\tau}(M_{SUSY})
\ , \nonumber
\end{eqnarray}
where, the variables with a prime$(')$ mean the redefinition 
of them in the SM region to keep their continuity 
between the regions of the SM and SUSY. 
With such redefinitions, the graphs in this paper is continuous 
even if crossing on the $M_{SUSY}$ border. \\
The conditions at GUT energy-scale $M_{X}$ are described 
in the following subsections. 

\subsection{On the gauge coupling constants} 
\begin{figure}
\centering
\includegraphics[width=15cm]{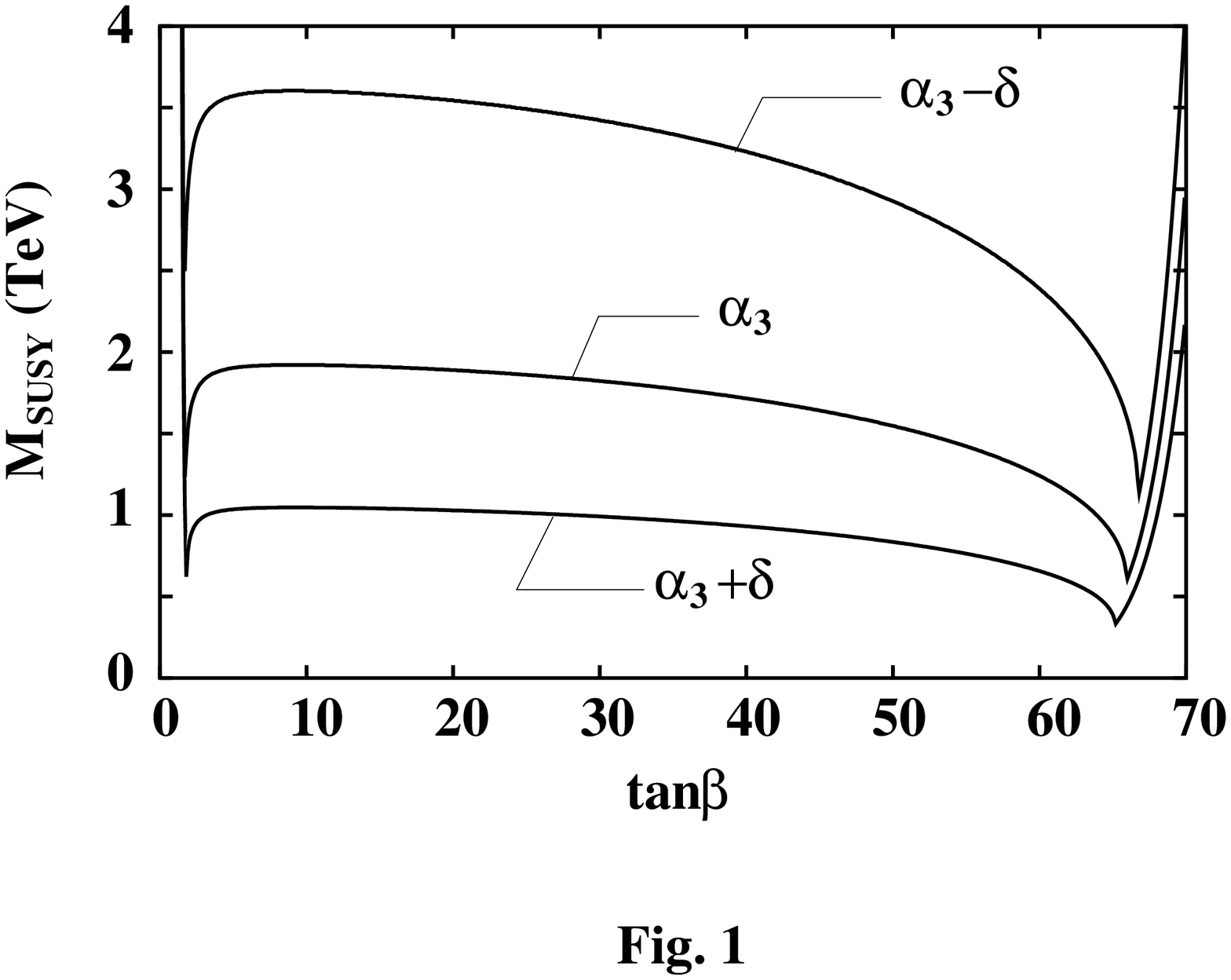}
\caption{\label{fig:msusy} 
$M_{SUSY}$ as functions of $\tan{\beta}$ 
for $\alpha_{3} \pm \delta = 0.1185 \pm 0.0020$. 
}
\end{figure}

The average value $\alpha_g$ among 
$\alpha_1, \alpha_2$, and $\alpha_3$ is introduced 
around the GUT scale $M_X$. 
$\chi^2_{g}$-fitting function is defined 
as the squared sum of each differences 
between $\alpha_g$ and $\alpha_i(i=1 \sim 3)$ as follows: 

\begin{equation}
\chi^2_{g}
\doteq
\sum_{i=1}^3\left(\frac{\alpha_i-\alpha_g}{\delta\alpha_i}\right)^2
\ , \label{eq:x2g} 
\end{equation}
where, $\delta\alpha_i$ means the uncertainty width of $\alpha_i$, 
respectively. the GUT scale $M_{X}$ 
is defined as the energy-scale where $\chi^{2}_{g}$ 
becomes the minimum. In order to make $\chi^2_{g}$ the minimum, 
$M_{SUSY}$ is fine-tuned. 
As the results, we find $\chi^2_{g} \simeq 0$ can be achieved 
by tuning $M_{SUSY}$ at any $\tan{\beta}$ 
in the ranges of 
the errorbars of the other parameters. 
Fine-tuned values of $M_{SUSY}$ are shown as functions of 
$\tan{\beta}$ with the uncertainty width of $\alpha_{3}$ 
in Fig.~\ref{fig:msusy} . 
We find the derived $M_{SUSY}$'s value depends on only 
the strong coupling constant $\alpha_3$ rather than 
the other parameters. 

\subsection{On the Yukawa coupling constants}
\begin{figure}
\centering
\includegraphics[width=15cm]{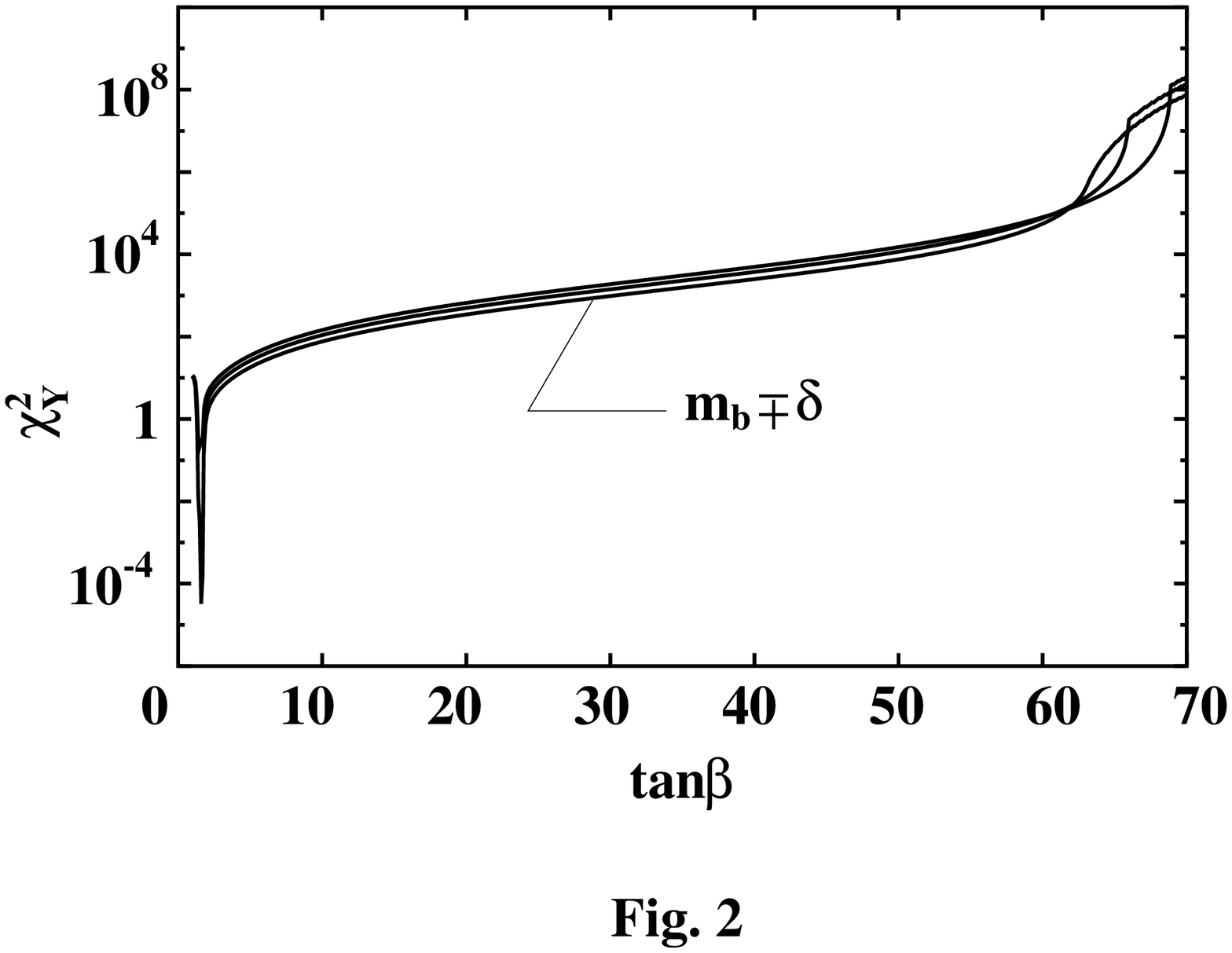}
\caption{\label{fig:x2ymbw} 
$\chi^2$ of the Yukawa coupling constants$(\chi^{2}_{Y})$ at 
GUT scale $M_{X}$ 
for $m_{b} \pm \delta = 4.2 \pm 0.2$(GeV) with $M_{SUSY}$. 
}
\end{figure}

When the SU(5) GUT scenario is presumed, 
Yukawa coupling constants of the bottom-quark and tau-lepton 
are unified from each other at $M_X$. 
Their $\chi^2_{Y}$-fitting function 
is defined as the following equations: 

\begin{equation}
\chi^2_{Y}
\doteq
\sum_{{q=b,\tau}}\left(
\frac{\alpha_{q}-\alpha_{Y}}{\delta\alpha_{q}}
\right)^2
\ , \label{eq:x2y} 
\end{equation}

\begin{equation}
\alpha_{Y}=\left(
\alpha_{b}+\alpha_{\tau}
\right)/2
\ , 
\end{equation}
where, $q$ implies $b$ or $\tau$, and $\delta\alpha_{q}$ means 
the uncertainty width of $\alpha_{q}$, respectively. 
With the uncertainty width $\delta = 0.2$(GeV) of $m_{b}$, 
$\chi^{2}_{Y}$ at $M_X$ scale as functions of $\tan{\beta}$ 
are shown in Fig.~\ref{fig:x2ymbw}. 
The behavior of $\chi^{2}_{Y}$ is not sensitive 
on the variance of $m_{t}$ or $\alpha_3$. 
Also shown in Fig.~\ref{fig:x2ymbw}, 
only small $\tan{\beta} \simeq 2$ is allowed 
by the likelihood analysis with the $\chi^{2}_{Y}$ function, 
and this minimal point of $\chi^{2}_{Y}$ is stable despite of 
the uncertainty of the parameters 
like $m_b$, $m_t$, or $\alpha_{3}$. 

\section{Analysis of $\tan{\beta}$ without $M_{SUSY}$}
\begin{figure}
\centering
\includegraphics[width=15cm]{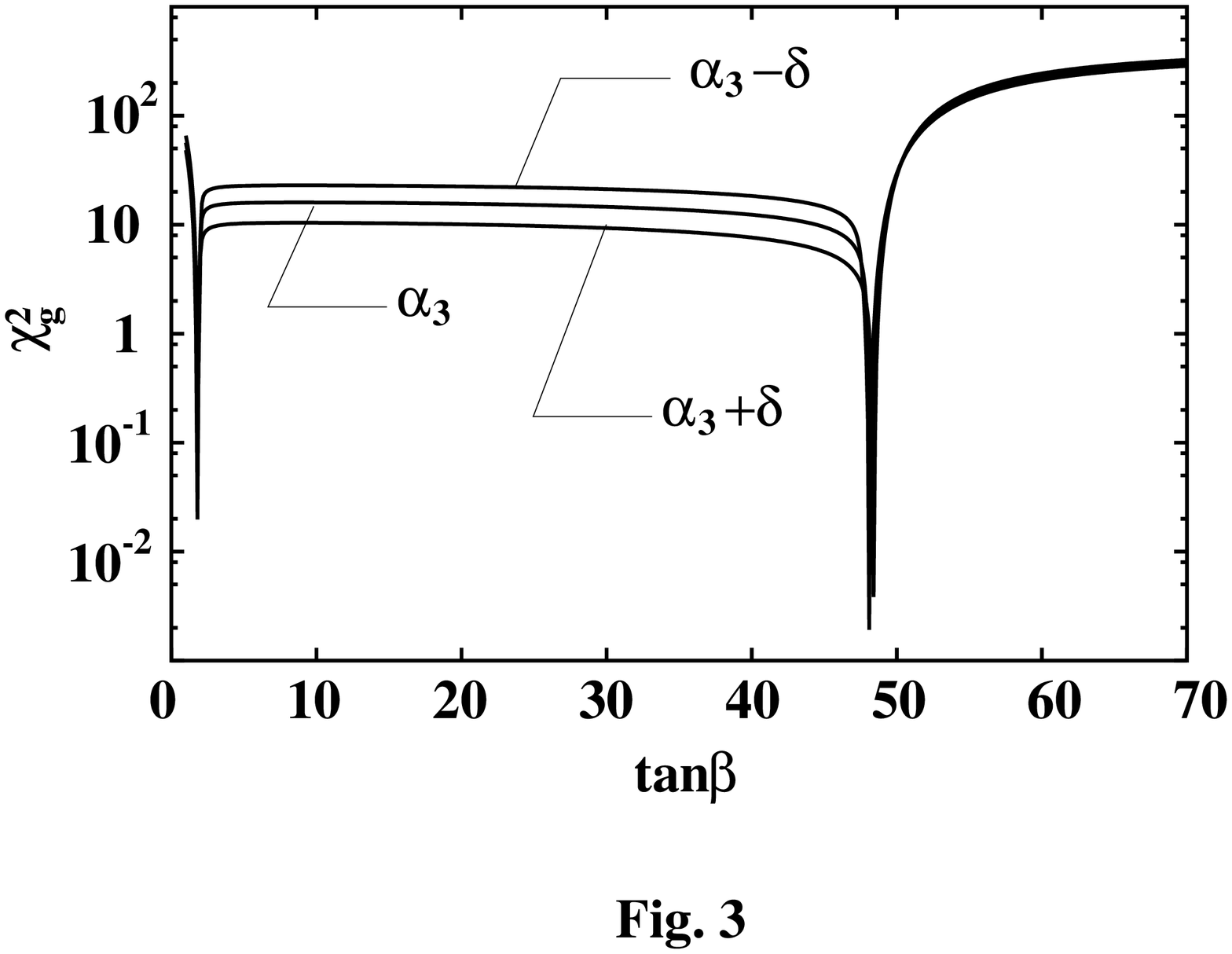}
\caption{\label{fig:x2ga3wo} 
$\chi^{2}$ of the gauge coupling constants$(\chi^{2}_{g})$ 
at $M_{X}$ for $\alpha_{3} \pm \delta = 0.1185 \pm 0.0020$ 
without $M_{SUSY}$. 
}
\end{figure}

\begin{figure}
\centering
\includegraphics[width=15cm]{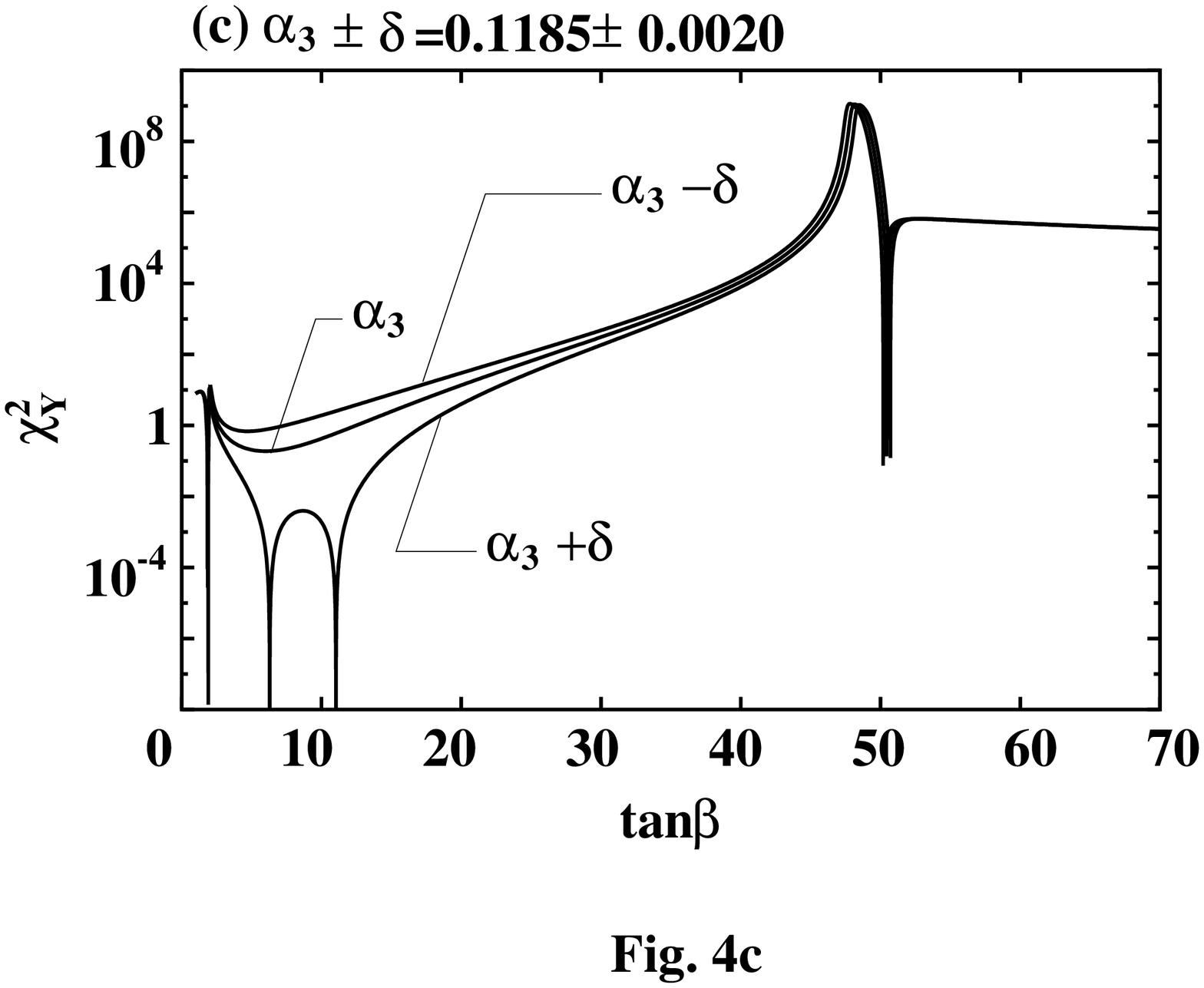}
\caption{\label{fig:x2ywo} $\chi^{2}_{Y}$ at $M_{X}$ 
for each uncertainty width of (a)$m_{b}$, (b)$m_{t}$, 
and (c)$\alpha_{3}$ without $M_{SUSY}$, respectively. 
}
\end{figure}

As mentioned above, on the model without $M_{SUSY}$ 
is discussed in this section. 
Therefore, the RGEs\cite{Castano} in this section are pure SUSY. 
On the boundary conditions of the gauge coupling constants, 
the VEVs of Higgs bosons, and $\tan{\beta}$, they are given 
at $M_{Z}$ in the same way of the model 
with including $M_{SUSY}$ as described in the previous section 
as from Eq.~(\ref{eq:ninva}) to Eq.~(\ref{eq:tanb}). 
The boundary conditions of the Yukawa coupling constants are 
given at the energy-scales of each on-shell masses as follows: 
\begin{equation}
\alpha_{t}(m_{t})
=
\frac{m_{t}^{2}(m_{t})}{2 \pi v^{2}(m_{t}) \sin^{2}\beta(m_{t})}
\ , 
\end{equation}
and 
\begin{equation}
\alpha_{b, \tau}(m_{b, \tau})
=
\frac{m_{b, \tau}^{2}(m_{b, \tau})}
{2 \pi v^{2}(m_{b, \tau}) \cos^{2}\beta(m_{b, \tau})}
\ . 
\end{equation}
The definitions of $\chi^{2}_{g}$ and $\chi^{2}_{Y}$ 
at $M_{X}$ are 
the same ones Eq.~(\ref{eq:x2g}) and Eq.~(\ref{eq:x2y}) 
in the previous section, however, in this section, 
$\chi^{2}_{g}$ depends on $\tan{\beta}$ or other parameters 
directly in contrast to the previous section, 
because the decoupling boundary $M_{SUSY}$ is 
vanished away from the model. \\
The dependence of $\tan{\beta}$ on $\chi^{2}_{g}$ with 
the uncertainty width 
of $\alpha_3$ is shown in Fig.~\ref{fig:x2ga3wo} . 
As Fig.~\ref{fig:x2ga3wo} shows, two minimal points exist 
at small and large $\tan{\beta}$. 
The variance of $\chi^{2}_{Y}$ at $M_{X}$ 
is shown with each uncertainty width of $m_b$, $m_{\tau}$, 
or $\alpha_3$ in Fig.~\ref{fig:x2ywo}a, \ref{fig:x2ywo}b, 
or \ref{fig:x2ywo}c, respectively. 
As shown in these graphs, 
usually $\chi^{2}_{Y}$ has three local minima. 

\section{Discussions} 
\subsection{On the model with $M_{SUSY}$}
\begin{figure}
\centering
\includegraphics[width=15cm]{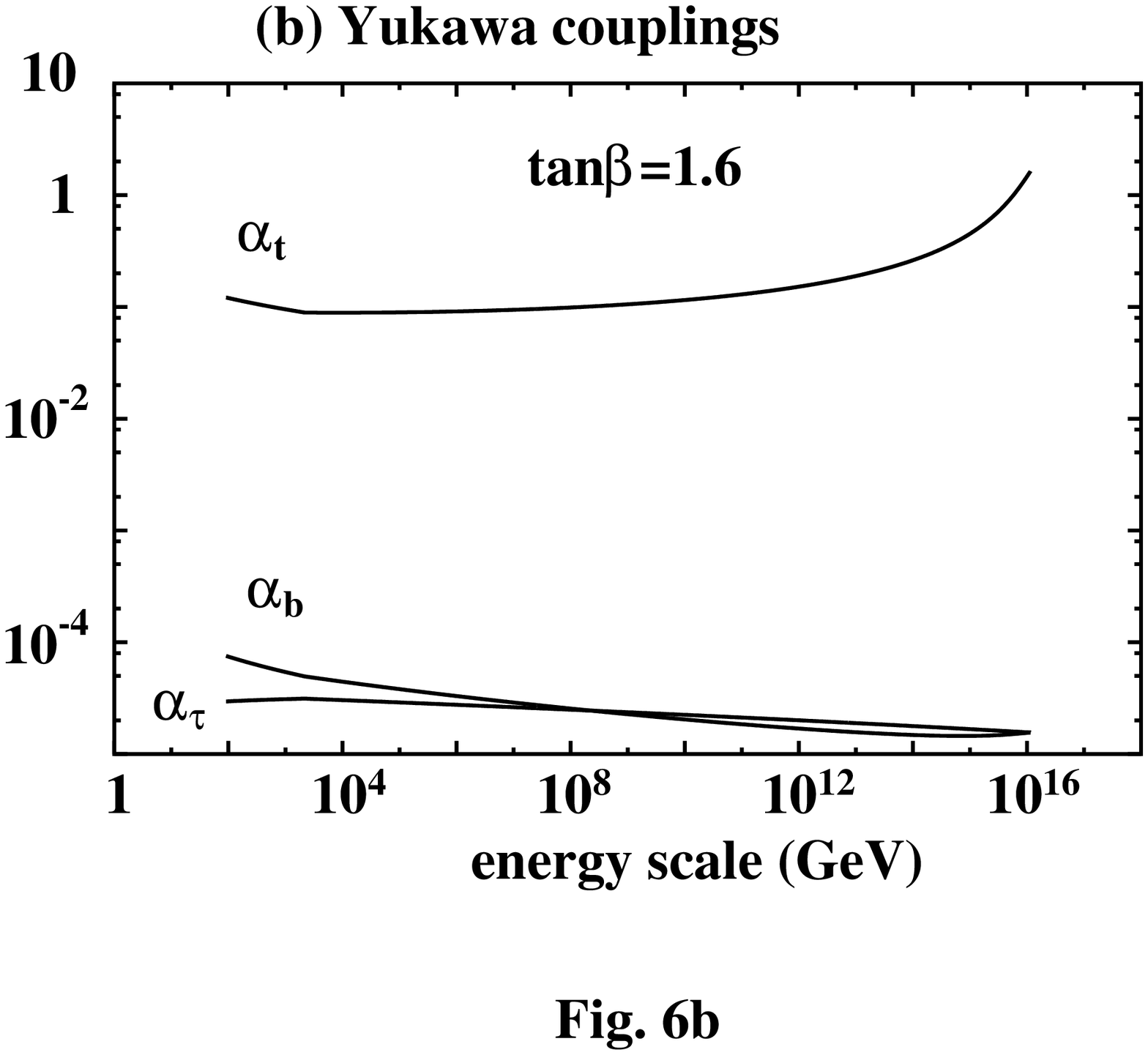}
\caption{\label{fig:gYc2w} 
The evolutions of the (a)gauge and (b)Yukawa coupling constants 
at $\tan{\beta}$=1.6 with $M_{SUSY}$. 
}
\end{figure}

Since $\chi^{2}$ of the gauge coupling constants($\chi^{2}_{g}$) 
can be always equal to zero by fine-tuning of $M_{SUSY}$, 
there exists no effective constraint on $\tan{\beta}$ 
from the unification of the gauge coupling constants. 
$\tan{\beta}$ is evaluated as 1.6 by the criteria of 
the bottom-tau unification in this model with $M_{SUSY}$. 
$M_{SUSY}$ is estimated about 2(TeV) 
to unify the gauge coupling constants at two-loop level. 
(As a reference, 
$M_{SUSY} \ll M_{Z}$ is derived at one-loop level. ) 
Nevertheless, such small $\tan{\beta} \simeq 2$ is difficult 
to be allowed by $g-2$ experiments\cite{Chattopadhyay}, 
because small $\tan{\beta}$ 
makes the masses of sparticles quite smaller than expected. 
The evolutions of the gauge and 
Yukawa coupling constants at $\tan{\beta}=1.6$ are shown 
in Fig.~\ref{fig:gYc2w}a and \ref{fig:gYc2w}b, respectively. 
The summary of the arguments in this subsection 
is shown in the Table~\ref{tab:wmsusy}. 

\subsection{On the model without $M_{SUSY}$}
\begin{figure}
\centering
\includegraphics[width=15cm]{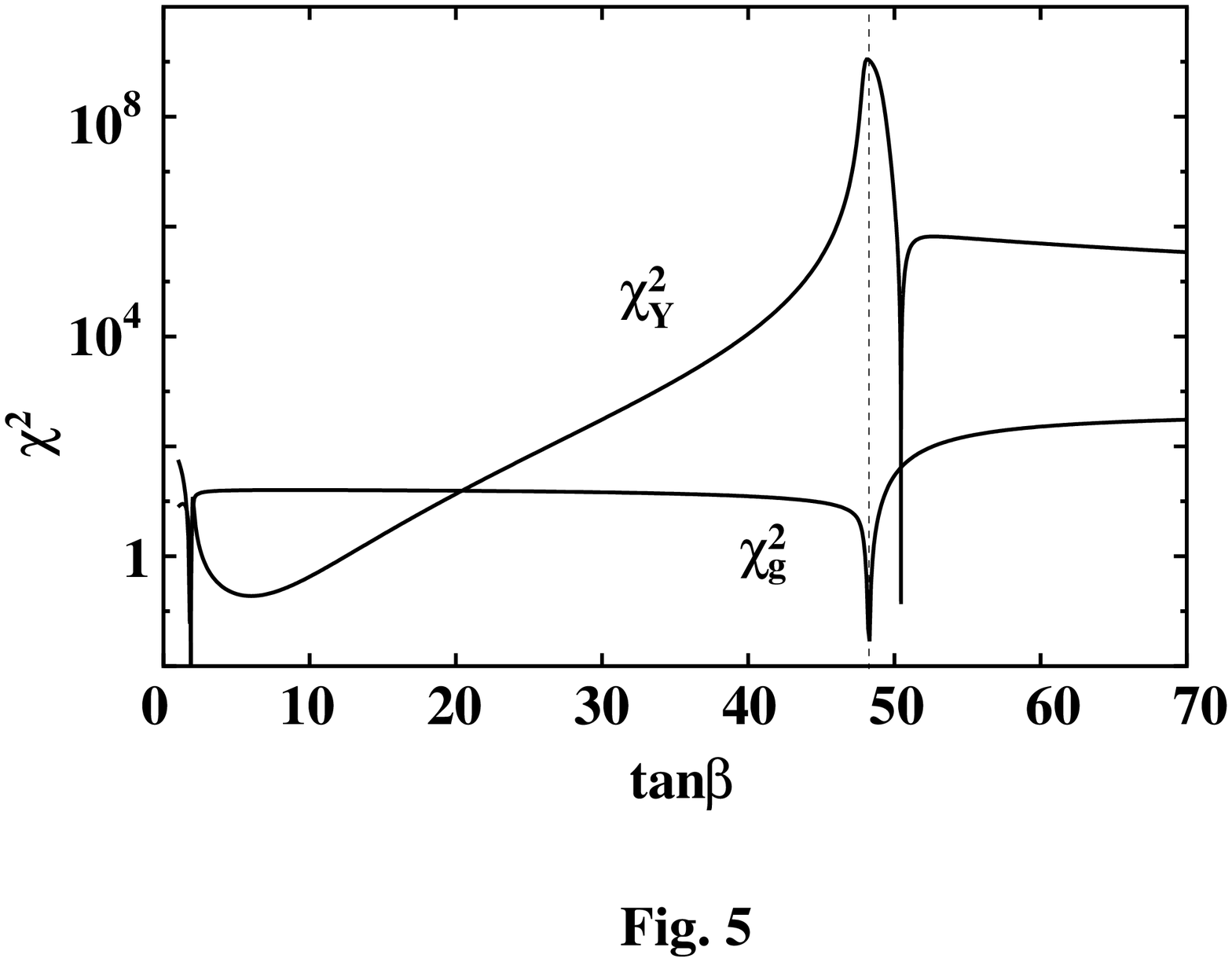}
\caption{\label{fig:x2yg50} 
Opposite behavior of $\chi^{2}_{Y}$ vs $\chi^{2}_{g}$
around large $\tan{\beta} \simeq 50$ 
at $M_{X}$ without $M_{SUSY}$. 
}
\end{figure}

\begin{figure}
\centering
\includegraphics[width=15cm]{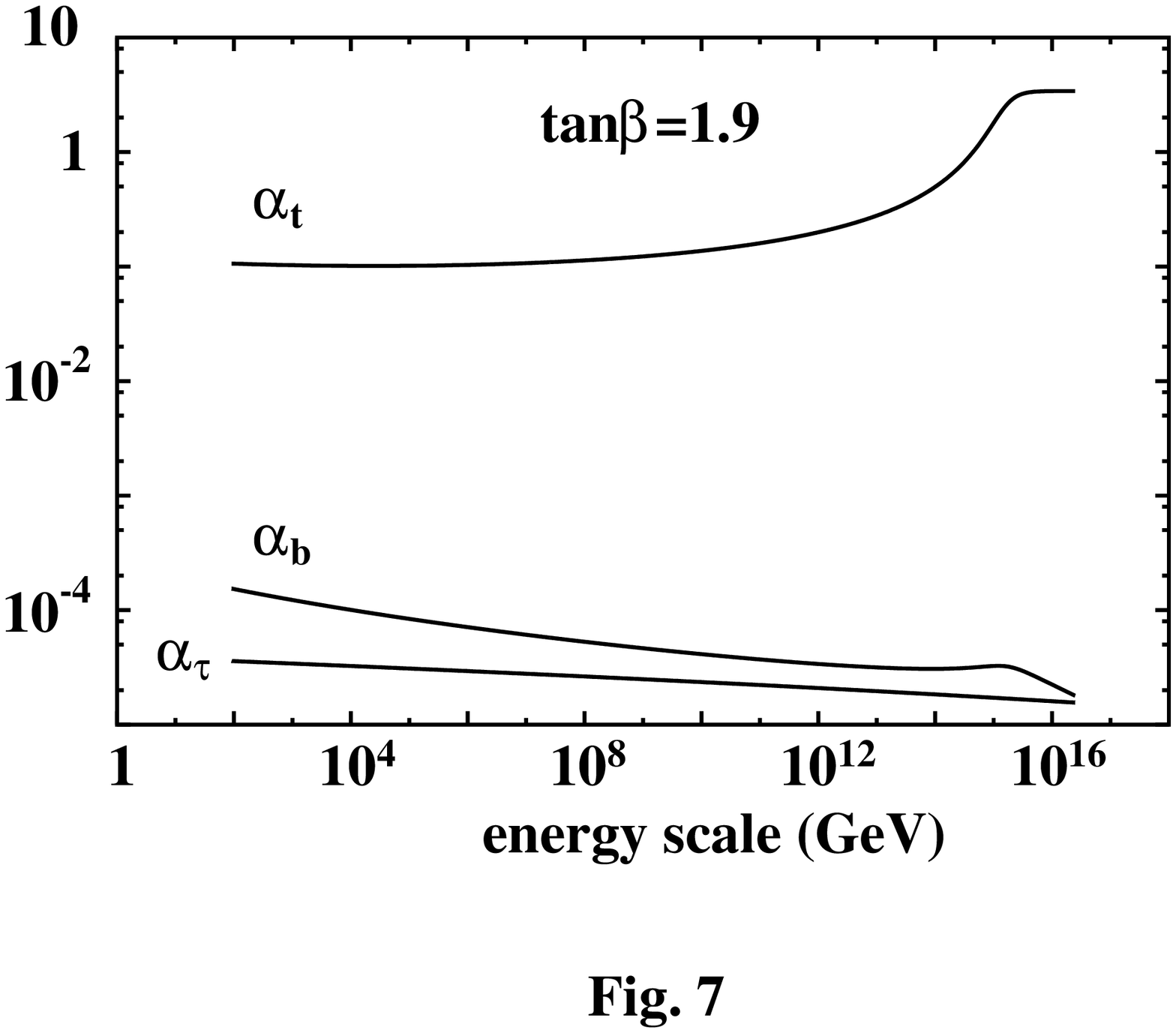}
\caption{\label{fig:Yc2wo} 
The evolution of the Yukawa coupling constants 
at $\tan{\beta}$=1.9 without $M_{SUSY}$. 
}
\end{figure}

\begin{figure}
\centering
\includegraphics[width=15cm]{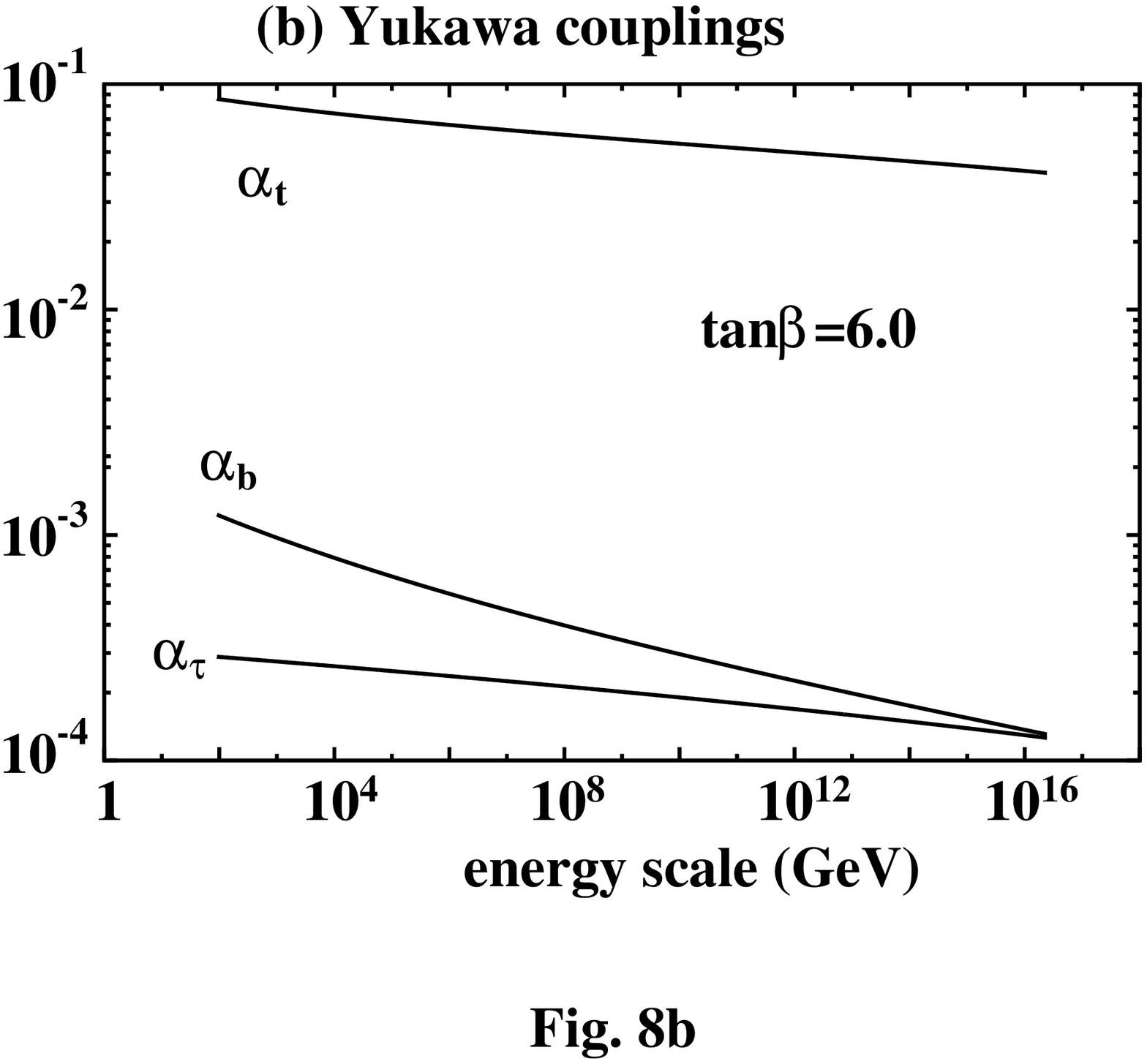}
\caption{\label{fig:gYc6wo} 
The evolutions of the (a)gauge and (b)Yukawa coupling constants 
at $\tan{\beta}$=6.0 without $M_{SUSY}$. 
}
\end{figure}

The value of $\tan{\beta}$ is possibly settled 
at 2, $6 \sim 25$, or 50 
in accordance with three minima of $\chi^{2}_{Y}$. 
However, large $\tan{\beta} \simeq50$ is denied theoretically, 
because $\chi^{2}_{g}$ and $\chi^{2}_{Y}$ cannot be minimal 
simultaneously, {\em i.e.}, they are exclusive from each other 
as shown in Fig.~\ref{fig:x2yg50}. 
On the small $\tan{\beta} \simeq 2$ choice, 
both of $\chi^{2}_{g}$ and $\chi^{2}_{Y}$ are 
enough small, however, this is excluded 
by the recent $g-2$ experiments\cite{Chattopadhyay} 
as mentioned previously. 
Large $\tan{\beta} \simeq 50$ is also not probable 
if $b \rightarrow s\gamma$ experiment\cite{b-sr} is true. 
Moreover, the authors are dubious whether 
these two local minima of $\chi^{2}_{Y}$ 
at small and large $\tan{\beta}$ 
are physical ones or not, because they make 
the RGEs divergent at one-loop level, 
therefore, the evolution of the Yukawa coupling constants 
seems strange even at two-loop level, for instance, as shown 
in Fig.~\ref{fig:Yc2wo} at $\tan{\beta}=1.9$ . 
Medium $\tan{\beta}$ as $6 \sim 25$ 
is good for $\chi^{2}_{Y}$ of the bottom-tau unification, 
however, $\chi^{2}_{g}$ of the gauge unification cannot be 
enough small at two-loop level. 
Thus, precise unification of the gauge coupling constants 
cannot occur with medium $\tan{\beta}$ as $6 \sim 25$. 
The evolutions of the gauge or Yukawa coupling constants 
at $\tan{\beta}=6.0$ are shown 
in Fig.~\ref{fig:gYc6wo}a or \ref{fig:gYc6wo}b, respectively. 
These discussions are summarized in the Table~\ref{tab:womsusy}. 

\section{Summary and Conclusion}
In the SU(5) GUT scenario, $\tan{\beta}$ is evaluated 
by making use of the RGEs of the gauge 
and Yukawa coupling constants. 
Only small $\tan{\beta} \simeq 2$ is allowed in the model 
including $M_{SUSY}$, and $\tan{\beta}$ is estimated 
as 2 or 6 to 25 in the model without $M_{SUSY}$, theoretically. 
However, small $\tan{\beta} \simeq 2$ is difficult to be allowed 
by the analysis of the recent $g-2$ experiment\cite{Chattopadhyay}. 
As the result, only medium $\tan{\beta}$ as $6 \sim 25$ 
is acceptable in the SUSY SU(5) GUT model without $M_{SUSY}$ 
if the bottom-tau unification is prior to the unification 
among the gauge coupling constants. 



\begin{acknowledgments}
The authors thank Y. Okada, M. Kakizaki and T. Nihei 
for useful advice. We are grateful to D. W. Hertzog for 
helpful comments on our preliminary stage. 
\end{acknowledgments}



\pagebreak

\begin{table}
\caption{\label{tab:wmsusy} 
Possiblity of $\tan{\beta}$ with $M_{SUSY}$. }
\begin{ruledtabular}
\begin{tabular}{lccc}
$\tan{\beta}$ 
& small$(\simeq 2)$ & medium$(6 \sim 25)$ & large$(\simeq 50)$ \\
\hline
$\chi^{2}_{g}$ & $\bigcirc$ & $\bigcirc$ & $\bigcirc$ \\
\hline
$\chi^{2}_{Y}$ & $\bigcirc$ & $\times$ & $\times$ \\
\hline
$g-2$ & $\times$ & $\bigcirc$ & $\bigcirc$ \\
\end{tabular}
\end{ruledtabular}
\end{table}
\begin{table}
\caption{\label{tab:womsusy} 
Possiblity of $\tan{\beta}$ without $M_{SUSY}$. }
\begin{ruledtabular}
\begin{tabular}{lccc}
$\tan{\beta}$ 
& small$(\simeq 2)$ & medium$(6 \sim 25)$ & large$(\simeq 50)$ \\
\hline
$\chi^{2}_{g}$ & $\bigcirc$ & $\bigtriangleup$ 
& $\bigcirc(\times)$\footnotemark[1] \\
\hline
$\chi^{2}_{Y}$ & $\bigcirc$ & $\bigcirc$ 
& $\times(\bigcirc)$\footnotemark[1] \\
\hline
$g-2$ & $\times$ & $\bigcirc$ & $\bigcirc$ \\
\hline
$b \rightarrow s \gamma$ 
& $\bigcirc$ & $\bigcirc$ & $\bigtriangleup$ \\
\end{tabular}
\end{ruledtabular}
\footnotetext[1]{These two scores are exclusive 
from each other.}
\end{table}


\pagebreak

%

\begin{figure}
\centering
\includegraphics[width=15cm]{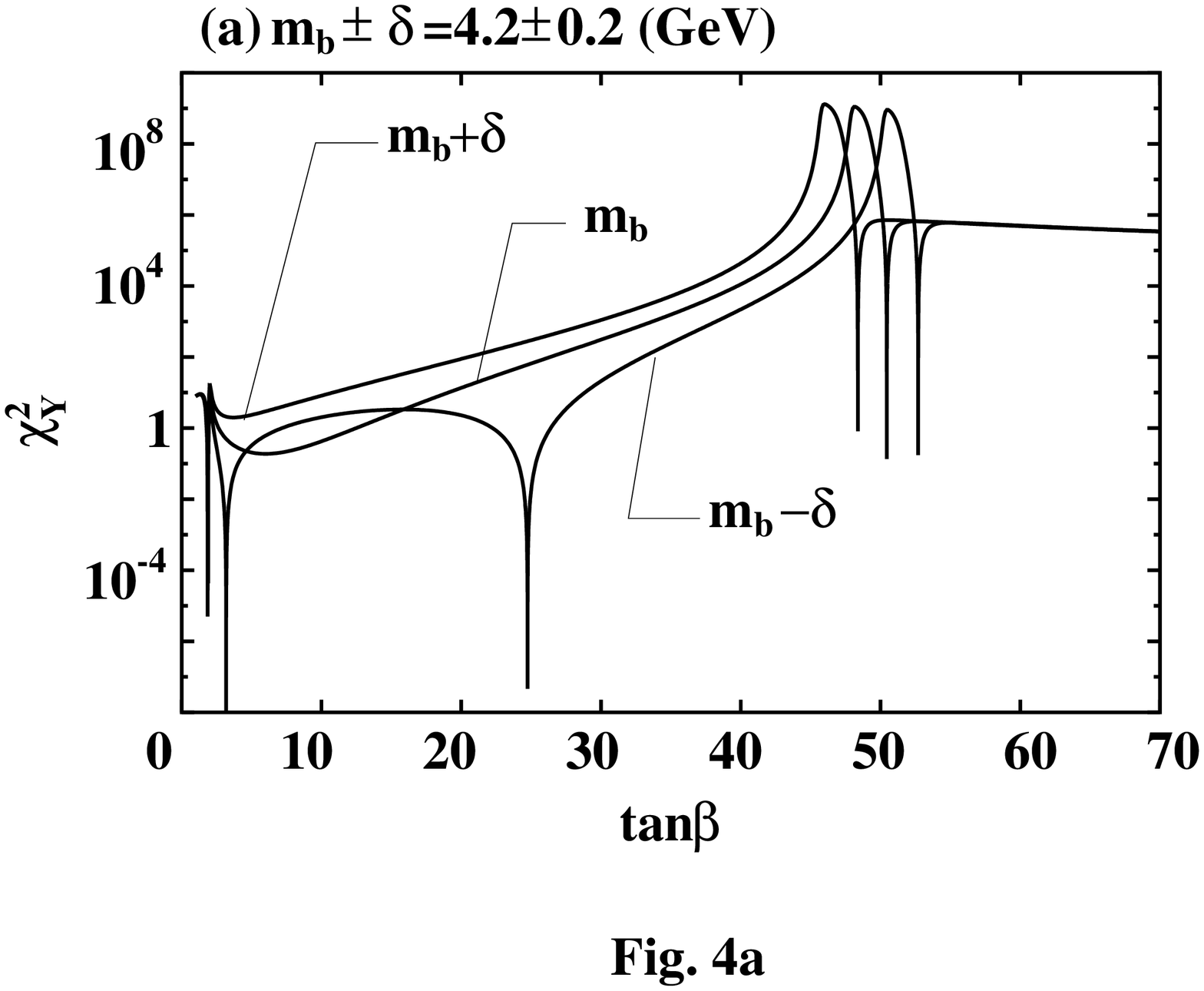}
\end{figure}
\begin{figure}
\centering
\includegraphics[width=15cm]{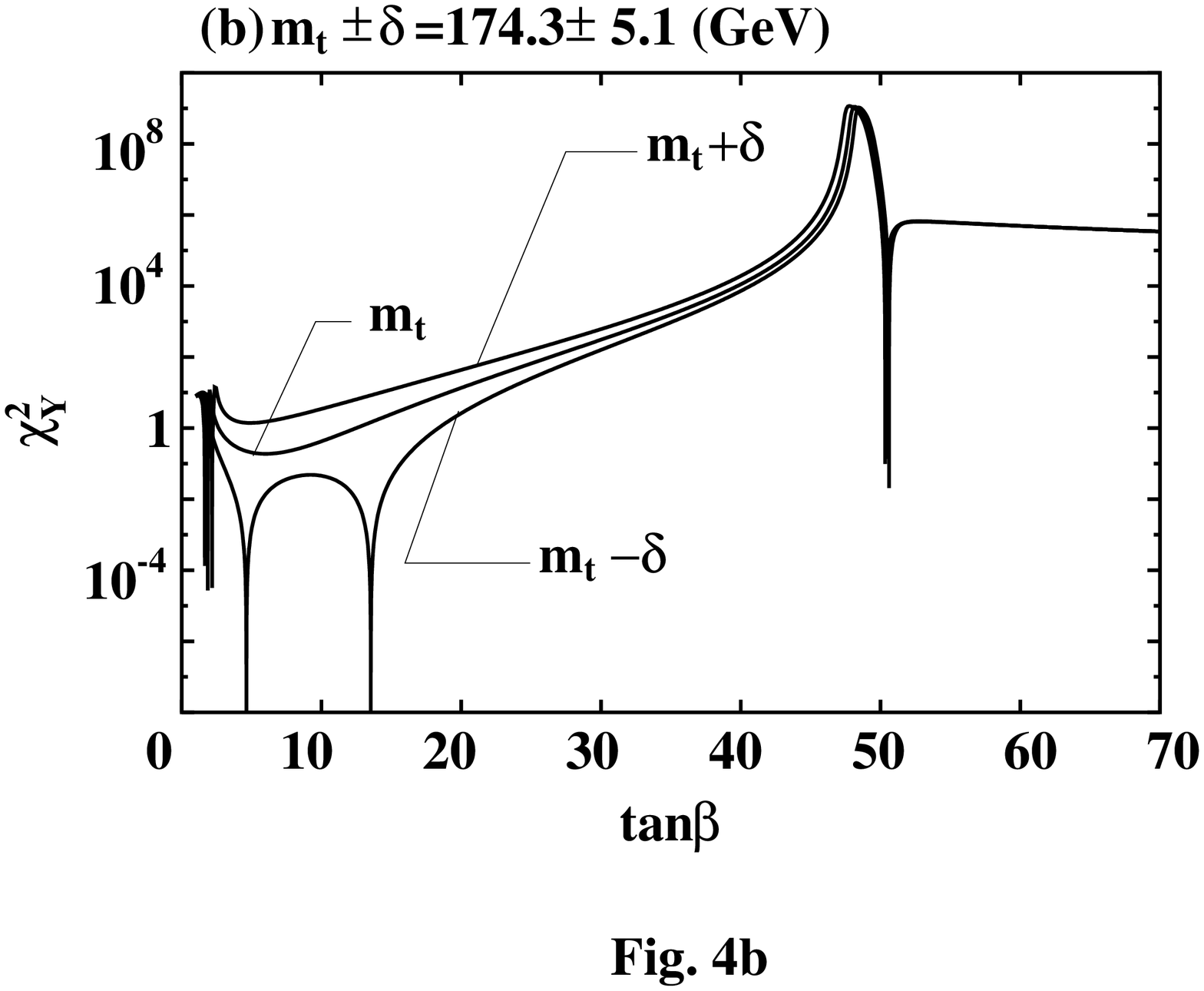}
\end{figure}
\begin{figure}
\centering
\includegraphics[width=15cm]{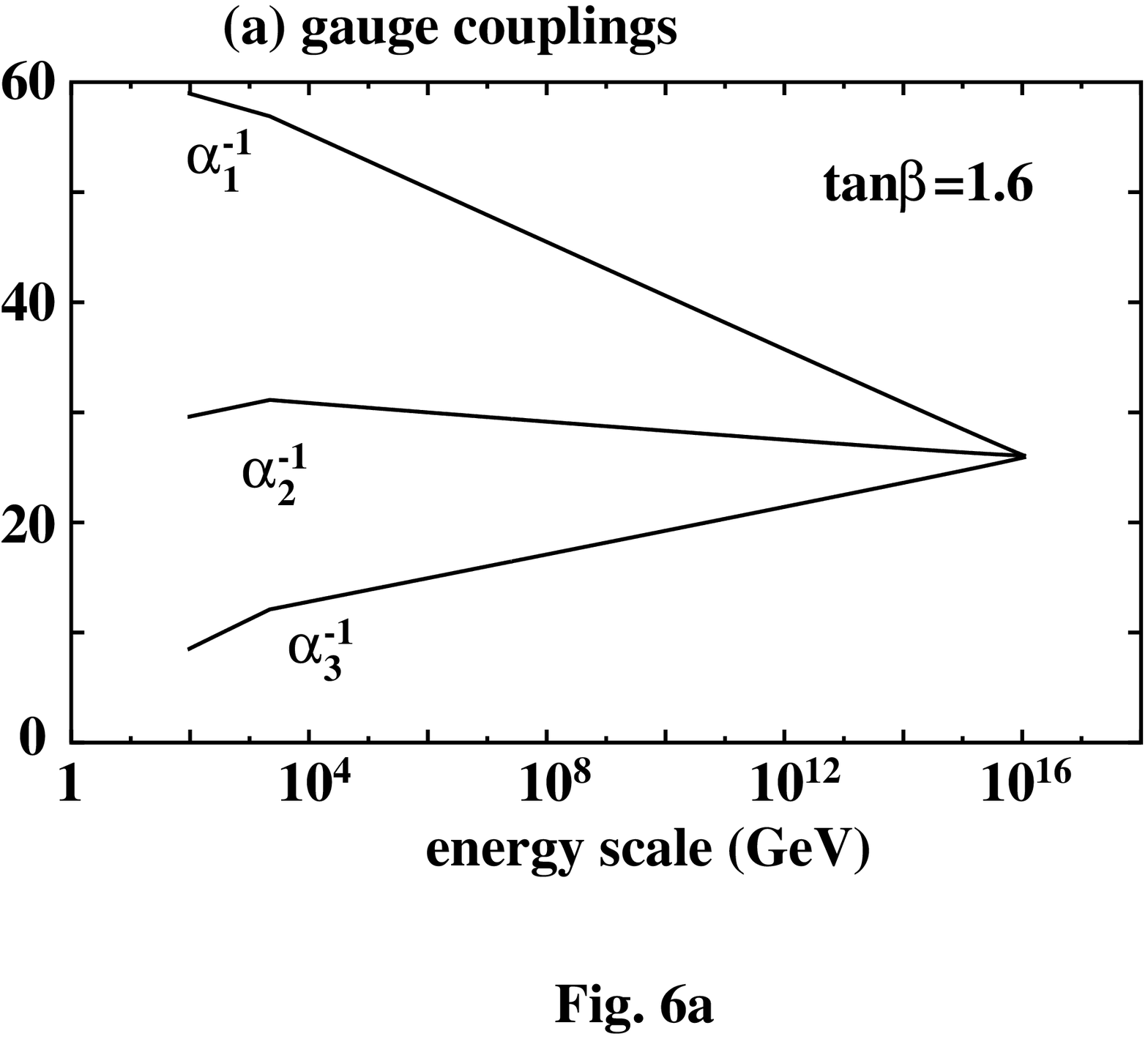}
\end{figure}
\begin{figure}
\centering
\includegraphics[width=15cm]{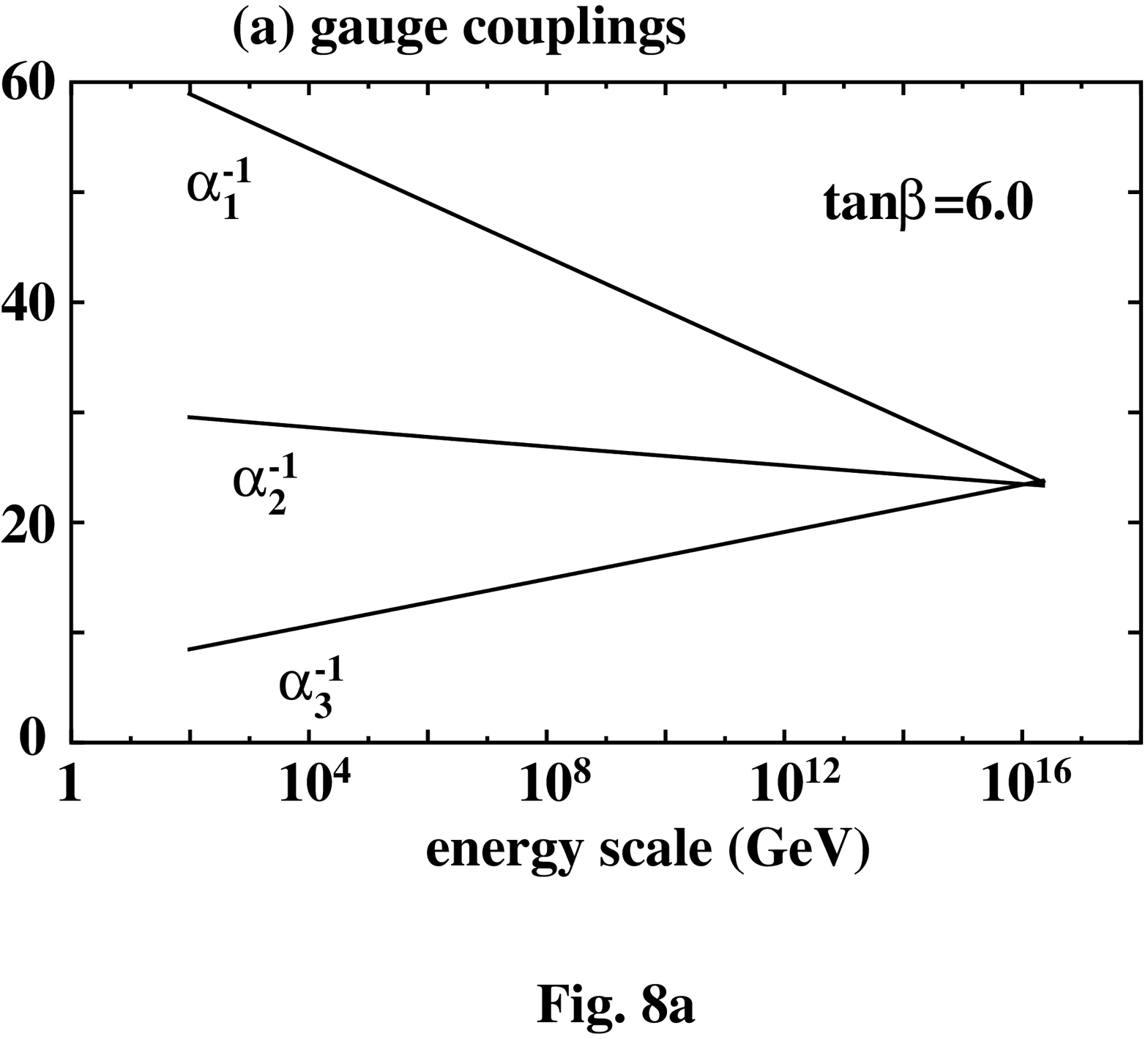}
\end{figure}


%

\end{document}